\RequirePackage{fix-cm}
\documentclass[smallextended]{svjour3}       % onecolumn (second format)
\smartqed  % flush right qed marks, e.g. at end of proof
\usepackage{graphicx}
\usepackage{natbib}
\usepackage{algorithm}
\usepackage[noend]{algpseudocode}
\usepackage{hyperref}
\begin{document}

\title{Pseudo-Outcrop Visualization of Borehole Images and Core Scans%\thanks{Grants or other notes
}


\author{Evgeny M. Mirkes \and Alexander N. Gorban \and  Jeremy Levesley \and Peter A. S. Elkington  \and  James A. Whetton}

\authorrunning{E.M. Mirkes et al.} % if too long for running head

\institute {E.M. Mirkes \at
University of Leicester, Department of Mathematics, Leicester, UK\\
\email{em322@le.ac.uk}\\
OrcID \url{http://orcid.org/0000-0003-1474-1734}
\and
A.N. Gorban \at 
University of Leicester, Department of Mathematics, Leicester, UK\\
\email{a.n.gorban@le.ac.uk}\\
OrcID \url{http://orcid.org/0000-0001-6224-1430}
\and
J. Levesley \at
University of Leicester, Department of Mathematics, Leicester, UK\\
\email{jl1@le.ac.uk}\\
OrcID \url{http://orcid.org/0000-0002-3509-0152}
\and
P.A.S. Elkington \at
Weatherford, Gotham Road, East Leake, Loughborough, LE12 6JX, UK\\
              Tel.: +44-115-9457880,
              Fax: +44-115-9405373\\
             \email{peter.elkington@eu.weatherford.com}\\
OrcID \url{http://orcid.org/0000-0001-8415-2260}
\and 
J. A. Whetton 
 \at
Weatherford, Gotham Road, East Leake, Loughborough, LE12 6JX, UK\\
OrcID \url{http://orcid.org/0000-0003-1527-9094}
}

\date{Received: date / Accepted: date}

\maketitle

\begin{abstract}
A pseudo-outcrop visualization is demonstrated for borehole and full-diameter rock core images to augment the ubiquitous unwrapped cylinder view and thereby to assist non-specialist interpreters. The pseudo-outcrop visualization is equivalent to a nonlinear projection of the image from borehole to earth frame of reference that creates a solid volume sliced longitudinally to reveal two or more faces in which the orientations of geological features indicate what is observed in the subsurface. A proxy for grain size is used to modulate the external dimensions of the plot to mimic profiles seen in real outcrops. The volume is created from a mixture of geological boundary elements and texture, the latter being the residue after the sum of boundary elements is subtracted from the original data. In the case of measurements from wireline microresistivity tools, whose circumferential coverage is substantially less than 100\%, the missing circumferential data is first inpainted using multiscale directional transforms, which decompose the image into its elemental building structures, before reconstructing the full image. The pseudo-outcrop view enables direct observation of the angular relationships between features and aids visual comparison between borehole and core images, especially for the interested non-specialist.
\end{abstract}

\section{Introduction}
\label{intro}
A common starting point for the geological analysis of circumferential images from boreholes and full-diameter cores is the identification of the types and orientations of bed boundaries and other discontinuities \citep{Rider2011}. Analysts determine the orientation of individual features and examine the ways in which the orientation of successive boundaries evolves over depth to delineate local structures and to infer depositional environments. In combination with other data at a variety of scales, image data helps refine geological models and helps define flow units and baffles in reservoir models. Image data also correlate with variations in mineralogy, porosity, and fluid content, making them useful in petrophysical analysis, \citep{xu2013,fu2016}. Additional textural information may be available from optical surface scans of full diameter rock cores; correlating these with borehole images in the same well allows the core and its features to be oriented in the earth's frame of reference via orientation data acquired by the logging tool.

High-resolution borehole images are routinely recorded, yet they are largely underutilized because of a lack of understanding of the data \citep{Joubert2016}. A contributing factor is the way in which the data are visualized; note that similar visualizations are applied to circumferential scans of full-diameter core. Both types of images are commonly displayed as unwrapped cylinders in which planar features manifest themselves as sinusoids, and appropriate software is needed to compute their individual orientations. The cylinder view has the advantage of being straightforward to generate, but it requires a mental transformation to visualize the angular relationships that would be observed in outcrops.

To observe boundary orientations in the earth's frame of reference, a solid volume corresponding to the material removed from the borehole must be constructed and sliced vertically in two planes. Modulating the external dimensions with a proxy for grain size gives a two-sided pseudo-outcrop view in which planar features are seen to be planar (rather than sinusoidal), and the angular relationships between them are observed directly without reference to other plots. All the information available to the reconstruction is located on the borehole wall or core surface; the interior has not been sampled. This differs from the typical three-dimensional (3D) inpainting problem in which the interior of an object is partially sampled. Aspect ratio is another differentiating characteristic because borehole images are long relative to their diameter. Image logs from a typical 0.2-m-diameter well commonly exceed 1,000 m in length; for the data sets examined here, the depth sample increment is 2 mm and comprises 176 circumferential samples. It is desirable to visualize data quickly.

\begin{figure}[!ht]
 \centering
  \includegraphics[width=0.9\textwidth]{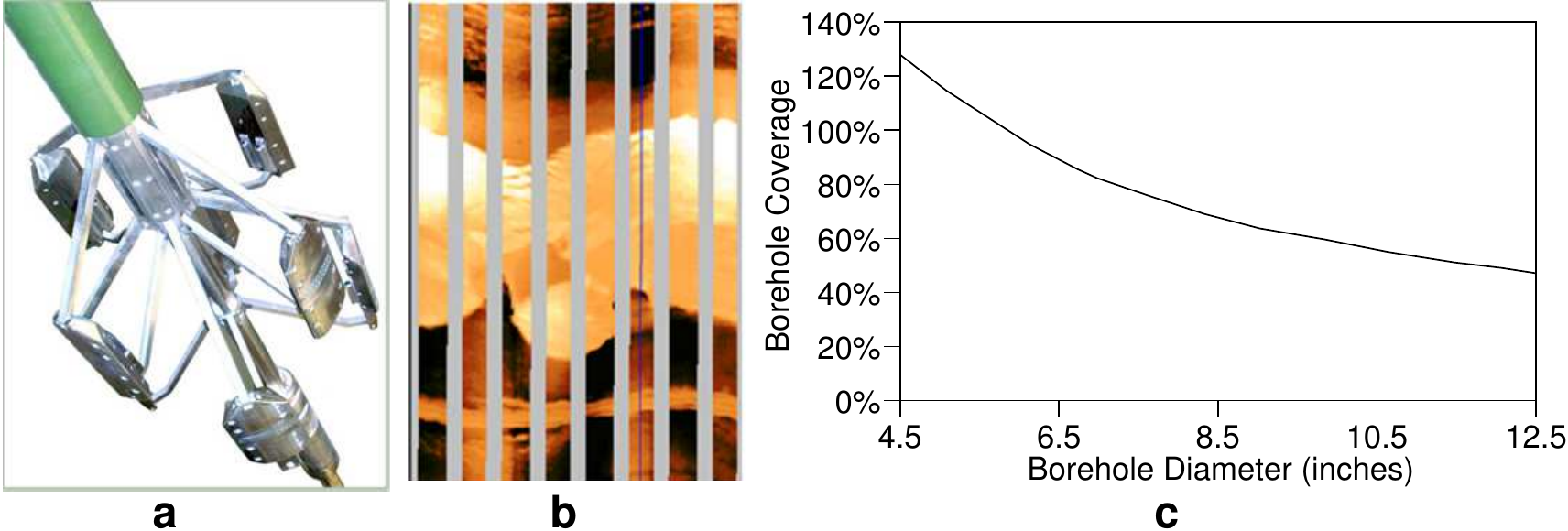}
\caption{ Microresitivity borehole imaging tool, pad section ({\bf a}), unwrapped image ({\bf b}), and circumferential coverage as a function of borehole diameter ({\bf c})}
\label{fig:1}       % Give a unique label
\end{figure}

In the case of wireline tool microresistivity measurements, current flows from individual 4-mm-diameter electrodes arranged in rows and inlaid in pads pressed against the borehole wall (Fig.~\ref{fig:1}(a)). In small-diameter wells, the rows overlap circumferentially to give full coverage; but when the well diameter exceeds approximately 150 mm (Fig.~\ref{fig:1}(c)), the space between pads creates longitudinal gaps in the image (Fig.~\ref{fig:1}(b)). The gaps are bound by linear margins in the tool's frame of reference.  Any 3D inpainting solution must handle the gaps directly;  alternatively, a two-dimensional (2D) inpainting solution must be applied as a preprocessing step.

In general there are two classes of algorithms for filling holes in digital images: texture synthesis and inpainting \citep{criminisi2004}. The former has been demonstrated for repeating patterns with some stochasticity and the latter for linear and curvilinear structures such as object contours. Both classes of algorithm have been applied to borehole images. \citet{hurley2011} use the statistical properties of the measured parts to generate realizations for filling the gaps in microresistivity images. This method can be appropriate for images dominated by textural features, such as vugs typical of carbonate rock formations; but achieving continuity across gap boundaries can be problematic, which makes it less well suited to images of rocks with complex curvilinear structures. This limitation is addressed by constraining the selection of realizations with structural information from locally adapted kernel regression, \citep{takeda2007, zhang2016}.  An inpainting method based on compressed sensing has been shown to successfully reconstruct full circumferential coverage in borehole images with up to 30\% coverage loss and up to 50\% in favourable cases \citep{assous2014}. The inpainting method builds on the general idea of minimizing the data required to represent information, which it seeks to do with the fewest number of multiscale directional transform coefficients such that reversing the process recovers the full data, including parts missing from the original samples \citep{elad2005}. The images are decomposed into elemental building structures (textures and piecewise smooth parts) that are combined to represent observed features \citep{starck2005}. This method is well suited to images containing bedding surfaces, fractures, slumps, clasts, and other curvilinear features typical of clastic rock formations and (depending on the choice of directional transform) is also able to handle textural elements.

Two-dimensional inpainting techniques have been extended to three dimensions for a variety of applications. In particular, multipoint geostatistics has been applied to filling gaps in micro-CT scans of porous rocks by extracting characteristics from a training image to generate a database of characteristics from which structures are selected to complete the missing parts in the target image \citep{zhang2014}. Statistical methods have also been applied to the creation of volumetric textures from 2D samples \citep{chen2010, urs2014}. The compressed sensing approach has also been extended to three dimensions; new multiscale directional transforms have been designed that enable capturing the true geometry of 3Dl objects (avoiding the risk of misrepresentation from slice-by-slice processing) \citep{woiselle2011}.

An alternative method is designed to be useful for images in which the primary structural elements are planar or subplanar at the borehole/core scale. This is the case in many hydrocarbon-bearing siliciclastic rocks logged by microresistivity imaging tools. The method will propagate solid textures but not necessarily in an optimal way. It can be easily combined with any advanced method of texture analysis if textural elements are the primary interest (such as in vuggy carbonates). It is tested on several real-life examples.

This  paper comprises three sections: Description of the method, Results, and Conclusion.

\section{Description of the Method}
\label{sec:1}
The principal steps in constructing the 3Dl volume are as follows
\begin{enumerate}
  \item Divide the log into overlapping depth windows. In each window, identify the dominant set of planar/subplanar geological features and remove them from the image. This is Level 1 decomposition.
  \item Identify the dominant planar feature set within the residual image and subtract it. This is Level 2 decomposition. Repeat the process several times until all of the subsidiary planar geological features are identified and the final residual image contains only non-planar features. The image may include strongly non-planar features and a host of other features that are localized azimuthally and will be referred to as residual texture.
  \item Identify functions to represent the shape of the primary geological feature and each subsidiary feature for all the levels of decomposition, and use these to construct a synthesised volume.
  \item Propagate the residual textures through the volume.
\end{enumerate}

Once created, the volume is visualized by, for example, making two longitudinal cuts through the oriented volume to expose a wedge of material, which allows examination of the orientations of geological features and their relationships in the earth's frame of reference. Additional information from the image measurement can be encoded as variations in the external dimensions of the wedge.

\begin{figure}
 \centering
  \includegraphics[width=0.5\textwidth]{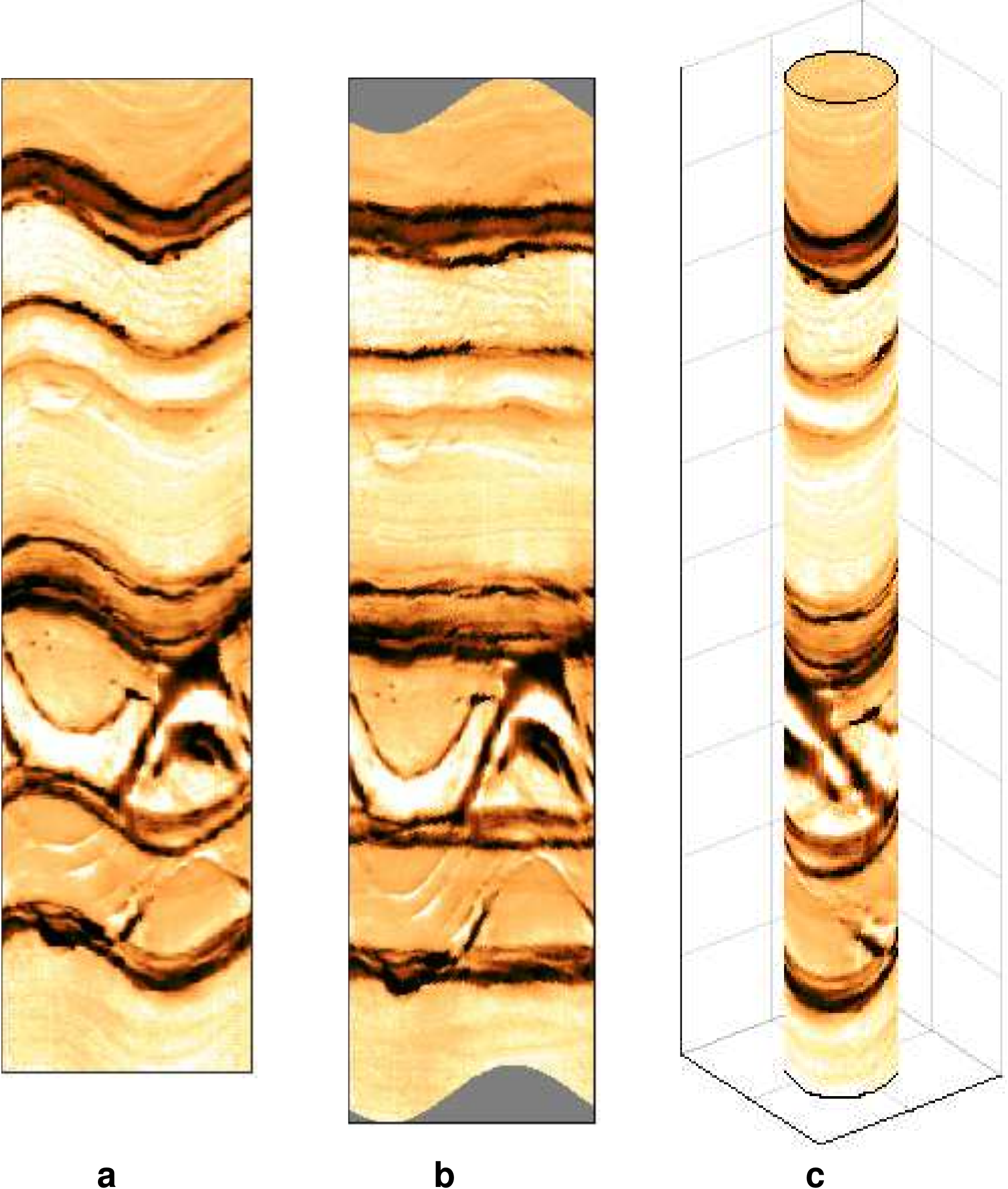}
\caption{The standard 2D borehole image ({\bf a}), 2D image after subtraction of the dominant dip sinusoid from the ordinate ({\bf b}), and cylindrical borehole surface in 3D coordinate systems ({\bf c})}
\label{fig:Image}       % Give a unique label
\end{figure}

Image of a borehole wall in several coordinate systems are presented in Fig.~\ref{fig:Image}. Let us consider one of the overlapping depth windows as a cylinder (Fig.~\ref{fig:Image}(c)). Before the earth frame transformation begins, the raw image-log data is processed in the usual way and stored in an array $f_{ij}$, where $i=1,\ldots,H$ are rows corresponding to depth and $j = 1,\ldots,N$ are columns corresponding to circumferential location (Fig.~\ref{fig:Image}(a)). These 2D arrays correspond to the surface of a cylinder (Fig.~\ref{fig:Image}(c)). The borehole has a nominal radius $R$ and the step between rows is $h$. Note that the data are discrete with step $h$ in the vertical direction, and $f_{ij}$ is the restriction of a scalar field $f$ to the surface of the borehole. Each position $(i,j)$ in the array $f_{ij}$ corresponds to the space radius-vector $\vec{r}_{ij}$  from the origin $O$  to the  point on the cylinder surface.

Some of the $f_{ij}$ may correspond to missing or bad measurements, so  a mask $m_{ij}$ is used  to identify the locations of valid values in which a value of 1 is used if a given $f_{ij}$ is known and 0 is used if $f_{ij}$ is unknown. At this point it may be advantageous to construct a full circumferential coverage image by application of the morphological components analysis method \citep{assous2014}.

\begin{figure}
 \centering
  \includegraphics[width=0.7\textwidth]{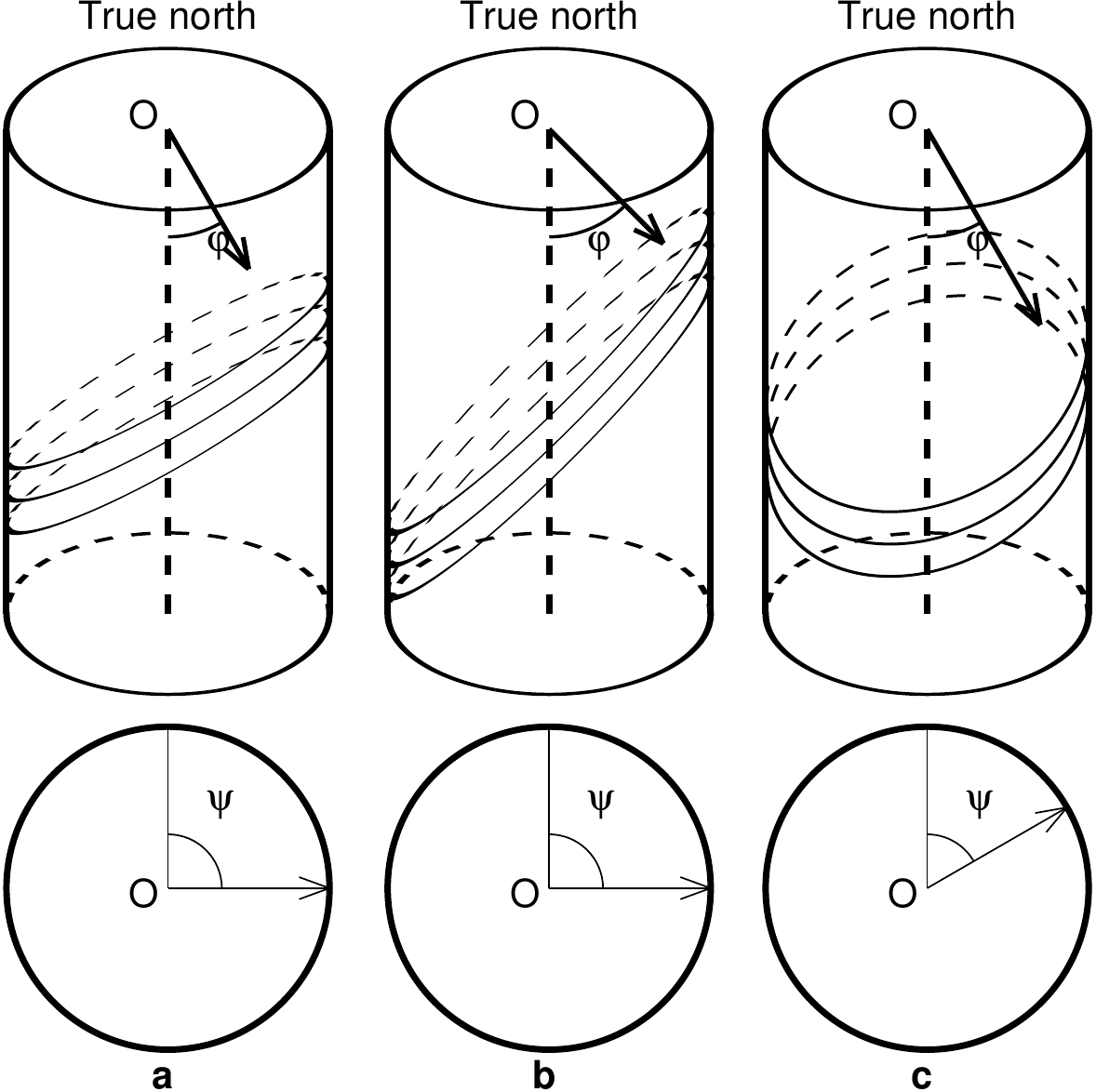}
\caption{Reference slope and orientation of planes ({\bf a}), steeper planes with increased $\phi$ ({\bf b}), and rotated planes with decreased $\psi$ ({\bf c})}
\label{fig:Planes}       % Give a unique label
\end{figure}

The planar structure shown in Fig.~\ref{fig:Planes} is an important part of the method. Two angles,  $\phi$ and $\psi$, describe the planes relative to a reference direction (True North in this case), and a distance coordinate identifies the stack of planes by sequence number.
 The fragment of the borehole between two neighboring planes is called a {\em slice}. The polar angle $\phi\in[0,\pi/2]$ is the angle that the normal vector $\vec{n}$ to the planes makes with the vertical axis. The azimuthal angle $\psi\in[0,2\pi]$ is the angle that the projection of the normal onto the base of the cylinder makes with True North.  For the degenerate case of zero $\phi$, the value of $\psi$ is set to zero. Three examples   illustrate the idea in Fig.~\ref{fig:Planes}  with Fig.~\ref{fig:Planes}(a) being a reference configuration. In Fig.~\ref{fig:Planes}(b)  an increase in $\phi$ coresponds to  a higher angle of dip, and in  (Fig.~\ref{fig:Planes}(c)) the angle $\psi$ has decreased  such that the planes have the same dip but the high point of the planes is rotated to a different position. All slices that have substantially the same angles $\phi$ and $\psi$ are designated as a slice family $S_{\phi\psi}$.

\begin{figure}
 \centering
  \includegraphics[width=0.9\textwidth]{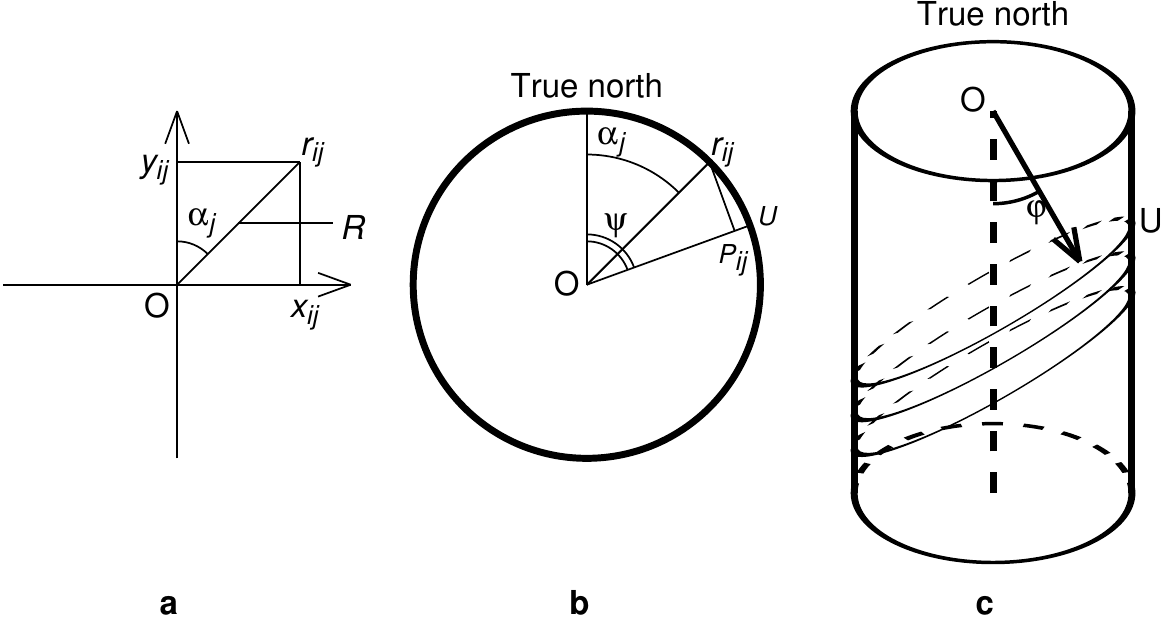}
\caption{Cartesian co-ordinate system ({\bf a}),  True North reference ({\bf b}), and planar slices ({\bf c})}
\label{fig:coordinate}       % Give a unique label
\end{figure}

In the Cartesian coordinate system (Fig.~\ref{fig:coordinate}), the origin $O$ is located at the centre of the upper base of a cylinder, the $y$-axis is directed towards True North, and the $x$ and $y$ coordinates of the point $f_{ij}$ are defined as
\begin{equation}
x_{ij}=R\sin \alpha_j, y_{ij}=R\cos⁡ \alpha_j,
\end{equation}
where $\alpha_j$ is the angle between the $y$-axis and the point $f_{ij}$ for the $j$ column and is defined as
\begin{equation}
\alpha_j=\frac{2\pi j}{N}.
\end{equation}

The $z$-axis is directed downward, and the $z$-coordinate of the point $f_{ij}$ is given by
\begin{equation}
z_{ij}=h(i-1).
\end{equation}
Assuming that all slices in the family have the same constant thickness equal to the spacing $h$ between adjacent members of the slice family, it is possible to enumerate the slices in the slice family by distance from the origin $O$. The Cartesian form of the unit normal vector \vec{n} is
\begin{equation}
x_n=\sin\phi\sin\psi,y_n=\sin\phi\cos\psi,z_n=\cos\phi.
\end{equation}

 The directed length of the projection of the vector $\vec{r}_{ij}$ on the unit normal vector is the dot product of the unit normal vector $\vec{n}$ and the vector $\vec{r}_{ij}$

\begin{equation}
p_{ij}=\vec{n}\cdot\vec{r}_{ij}=R\sin\phi\cos⁡(\psi-\alpha_j)+h(i-1)\cos\phi.
\end{equation}

The enumeration of the slice is then this length divided by the depth of one slice $h$
\begin{equation}
\nu_{ij}=\left[\frac{p_{ij}}{h}\right]=\left[\frac{R}{h}\sin\phi\cos⁡(\psi-\alpha_j)+(i-1)\cos⁡\phi\right],
\end{equation}
where $[\cdot]$ is rounding to the nearest integer. This describes the slice family and the number of the slice within the family.

All points in slice $p$ of an ideal slice family have the same value $\overline{f(p)}$. 
The best set of planes is computed by minimising  the variance of residual of the actual data after subtraction of the approximation
\begin{equation}
\label{eq:fit}
V(\phi,\psi) =var(f,S_{\phi\psi})=\frac{1}{\sum_{ij}m_{ij}}\sum_{m_{ij}=1}{\left(f_{ij}-\overline{f(\nu_{ij})}\right)^2}.
\end{equation}
It can be easily shown that the optimal slice constant $\overline{f(p)}$ is the mean of all points $f_{ij}$ of this slice
\begin{equation}
\overline{f(p)}=\frac{\sum_{\nu_{ij}=p, \, m_{ij}=1}{f_{ij}}}{\sum_{v_{ij}=p}{m_{ij}}}.
\end{equation}
	
\begin{figure}
 \centering
  \includegraphics[width=0.95\textwidth]{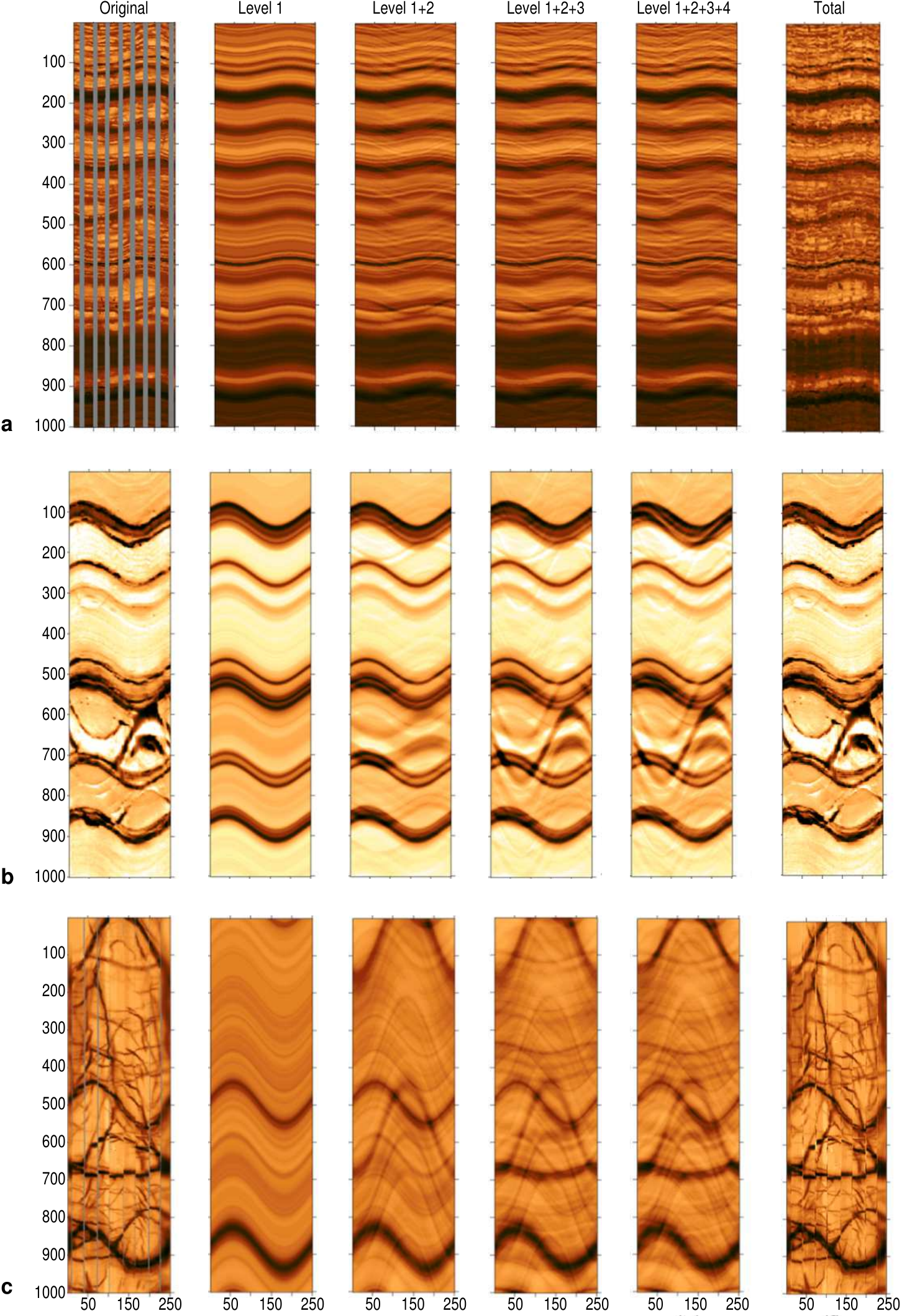}
\caption{Three images for the case study: Image with 65\% circumferential coverage ({\bf a}),   65\% circumferential coverage, two-dimensionally inpainted ({\bf b}), and 95\% coverage ({\bf c})}
\label{fig:Cases}       % Give a unique label
\end{figure}

\begin{figure}
 \centering
  \includegraphics[width=0.95\textwidth]{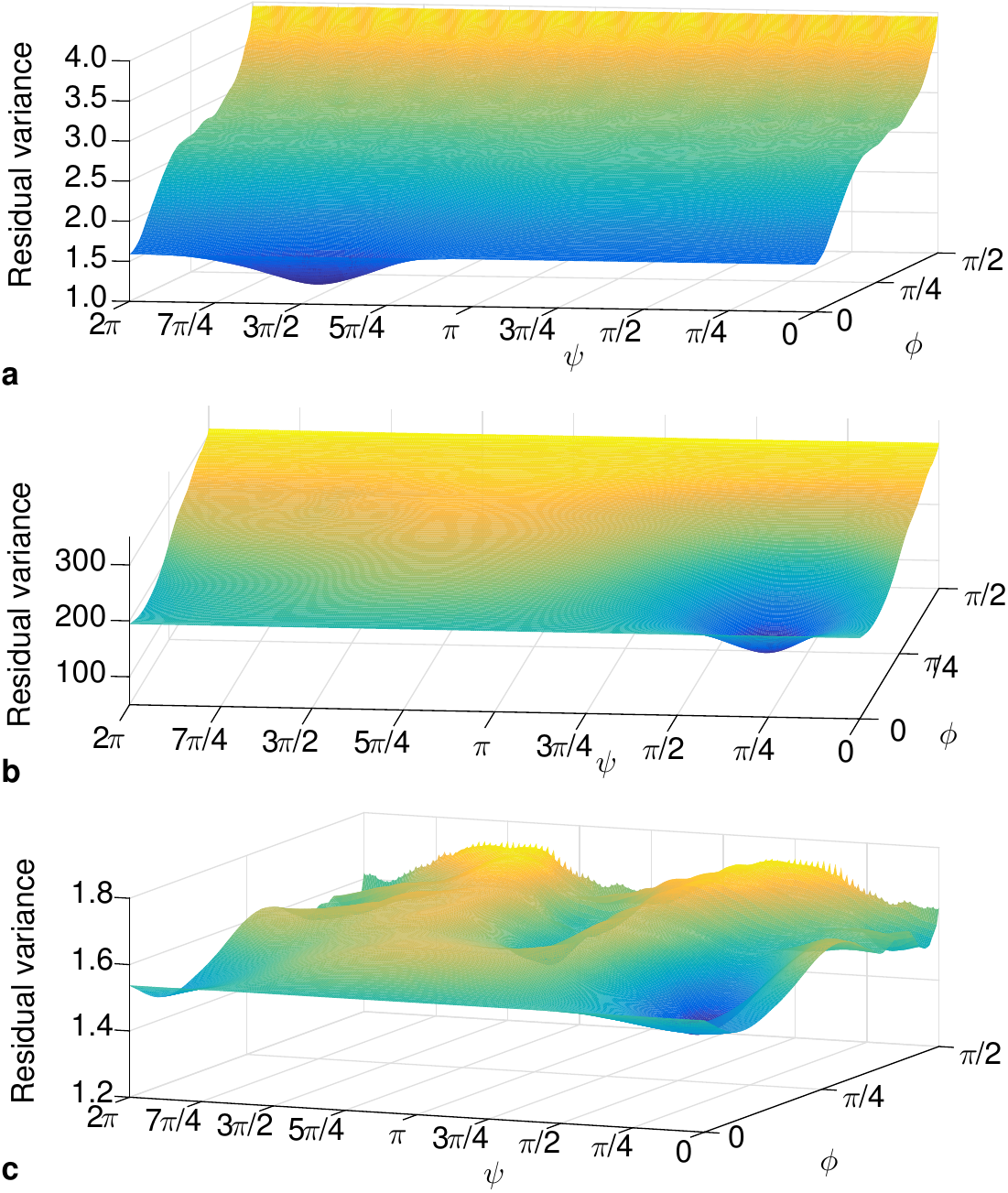}
\caption{Two-dimensional plots of the variance of residual (\ref{eq:fit}) for the dominant dip calculated for the borehole wall images Fig.~\ref{fig:Cases}(a), \ref{fig:Cases}(b), and \ref{fig:Cases}(c)   (in the same order)}
\label{fig:surf}       % Give a unique label
\end{figure}

Three images of real  borehole walls were selected for the case study (Fig.~\ref{fig:Cases}(a), \ref{fig:Cases}(b),  and \ref{fig:Cases}(c)).  The initial images, the dominant dip sets, the dominant dip sets plus 2nd, 3rd and 4th level dips, and sums of dip sets plus residual texture (the `total' column) are presented in Fig.~\ref{fig:Cases} for all cases. The model slice family is found by minimising the expression (\ref{eq:fit}) over all polar and azimuthal angles. Surfaces of this function, which depends on a particular choice of angles, are presented in Fig.~\ref{fig:surf}  for borehole walls from Fig.~\ref{fig:Cases}. The second (local) minimum on the bottom surface corresponds approximately to the second dip of the case Fig.~\ref{fig:Cases}(c). There appear no problems discretising the slices when they are parallel to the borehole because of convenient parameterisation. The Nelder-Mead algorithm (Algorithm \ref{alg1}) \citep{nelder1965} is used with several random starts to seek the minimum of the residual (\ref{eq:fit}). The random starts were repeated until the best solution appeared twice within specified accuracy. Practically, the global minimum of the residual variance has a large domain of attraction and minimisation is not significantly difficult or computationally expensive if the depths of the window exceeds its circumference (the width). If the aspect ratio approaches 1 (a rare situation) then the optimisation landscape becomes more noisy and a special smoothing is needed for optimisation. If the optimisation landscape is smooth (Fig.~\ref{fig:surf}) and the domain of attraction for the global minimum is large then the solution of the optimisation problem is expected to be robust with respect to small  perturbations.

\begin{algorithm}[!t]
\caption{Minimization of function $V(\phi,\psi)$ (7)}
\label{alg1}
\begin{algorithmic}[1]
    \State $\alpha\gets 1, \beta\gets 2, \gamma\gets 0.5, \sigma\gets 0.5$  \Comment Initiate parameters of the Nelder-Mead algorithm
    \State  $\psi_1\gets 0, \phi_1\gets 0, V^*\gets 2V(0,0),\varepsilon \gets 0.0001 $ \Comment Initiate search parameters
    \Repeat \Comment Search of the second case of the best solution within specified accuracy
        \If{$V^*>V(\psi_1,\phi_1)$}
            \State  $\psi^*\gets \psi_1, \phi^*\gets \phi_1, V^*\gets V(\psi_1,\phi_1)$ \Comment Save the best solution
        \EndIf
        \State $\psi_1\gets 2\pi r_1,\phi_1\gets \pi r_2/2$ \Comment Generate centre of simplex; $r_1,r_2$ are random numbers
        \State $\psi_2\gets \psi_1+\pi, \phi_2\gets\phi_1,\psi_3\gets\psi_1,\phi_3\gets\phi_1+\pi/4$ \Comment Complete simplex
    \Repeat \Comment the Nelder-Mead search for specified initial point
    \State Sort vertices to hold $V(\psi_1,\phi_1)\leq V(\psi_2,\phi_2)\leq V(\psi_3,\phi_3)$ \Comment Ordering
    \State $\psi_c\gets(\psi_1+\psi_2)/2, \phi_c\gets(\phi_1+\phi_2)/2$ \Comment Centroid calculation
    \State $\psi_r\gets \psi_c + \alpha (\psi_c - \psi_3),\phi_r\gets \phi_c + \alpha (\phi_c - \phi_3)$ \Comment Calculate reflection point
    \If {$V(\psi_1,\phi_1)\leq V(\psi_r,\phi_r) < V(\psi_2,\phi_2)$}
      \State $\psi_3\gets \psi_r,\phi_3\gets \phi_r$ \Comment Reflection
    \ElsIf {$(V(\psi_r,\phi_r)\leq V(\psi_1,\phi_1))$}
      \State $\psi_e\gets \psi_c + \beta(\psi_r - \psi_c ),\phi_e\gets \phi_c + \beta(\phi_r - \phi_c )$ \Comment Calculate expansion point
      \If {$(V(\psi_e,\phi_e)\leq  V(\psi_r,\phi_r))$}
        \State $\psi_3\gets \psi_e,\phi_3\gets \phi_e$ \Comment Expanding
      \Else
        \State $\psi_3\gets \psi_r,\phi_3\gets \phi_r$ \Comment Reflection
      \EndIf
    \Else \Comment Now $V(\psi_r,\phi_r)\geq V(\psi_2,\phi_2)$
      \State $\psi_{co}\gets \psi_c + \gamma(\psi_3 - \psi_c),\phi_{co}\gets \phi_c + \gamma(\phi_3 - \phi_c)$ \Comment Calculate contraction point
      \If {$V(\psi_{co},\phi_{co})\leq V(\psi_3,\phi_3)$}
        \State  $\psi_3\gets \psi_{co},\phi_3\gets \phi_{co}$ \Comment Contracting
      \Else
        \State $ \psi_j\gets \psi_1 + \sigma (\psi_j - \psi_1),\phi_j\gets \phi_1 + \sigma (\phi_j - \phi_1),   j = 2,3$ \Comment Shrinking
      \EndIf
    \EndIf
    \Until{$\min_{i,j} |\psi_i-\psi_j|\leq\varepsilon$ and $\min_{i,j} |\phi_i-\phi_j|\leq\varepsilon$}
    \Until{$|\psi^*-\psi_1|\leq\varepsilon$ and $|\phi^*+\phi_1|\leq\varepsilon$}
\end{algorithmic}
\end{algorithm}
 
After finding the best slice family  the residual field is computed by subtracting the values on the slice family found from the known field $f$
$$
res_{ij}^1=f_{ij}-s_{ij}^1,\forall m_{ij}=1,
$$
where $s_{ij}^1$ is the value of approximation of the field $f$ at the point $ij$ by the best slice family. This value is equal to $\overline{f(\nu_{ij})}$. The secondary slice structure $s^2$ is found by repeating the same procedure for the residual random field $r^1$ instead of $f$. Thus, recursively a sequence of layers $s^1,s^2,\ldots,s^L$ can be removed until the statistics of the residual are more or less random (or at a user-selected level), leaving behind a field that is considered to be  residual texture denoted by $t_{ij}$.

One value of the field is assigned per slice. This is  the mean of all values $f_{ij}$ of this slice
 $\overline{f(p)}$. This value is considered as a constant when extended in the modeled volume. For each $k=1,\ldots, L$  the slice $s^k$ is inpainted by this value inside the cylinder. These functions $s^k(x,y,z)$ are called below the {\em boundary elements}. The boundary element $s^{k+1}(x,y,z)$ provides the best approximation of the residual
 \begin{equation}\label{res^k}
 res_{ij}^k=f_{ij}-\sum_{q=1}^k s_{ij}^q
 \end{equation}
on the cylinder boundary by the function, which is constant in every slice from a system of parallel slices.  The approximation of the image is a sum of these boundary elements (\ref{res^k}). This sum is not constant in any slice for $k>1$ for a general situation when the original field $f_{ij}$ is not constant in a slice. 

\begin{figure}
 \centering
 \includegraphics[width=0.95\textwidth]{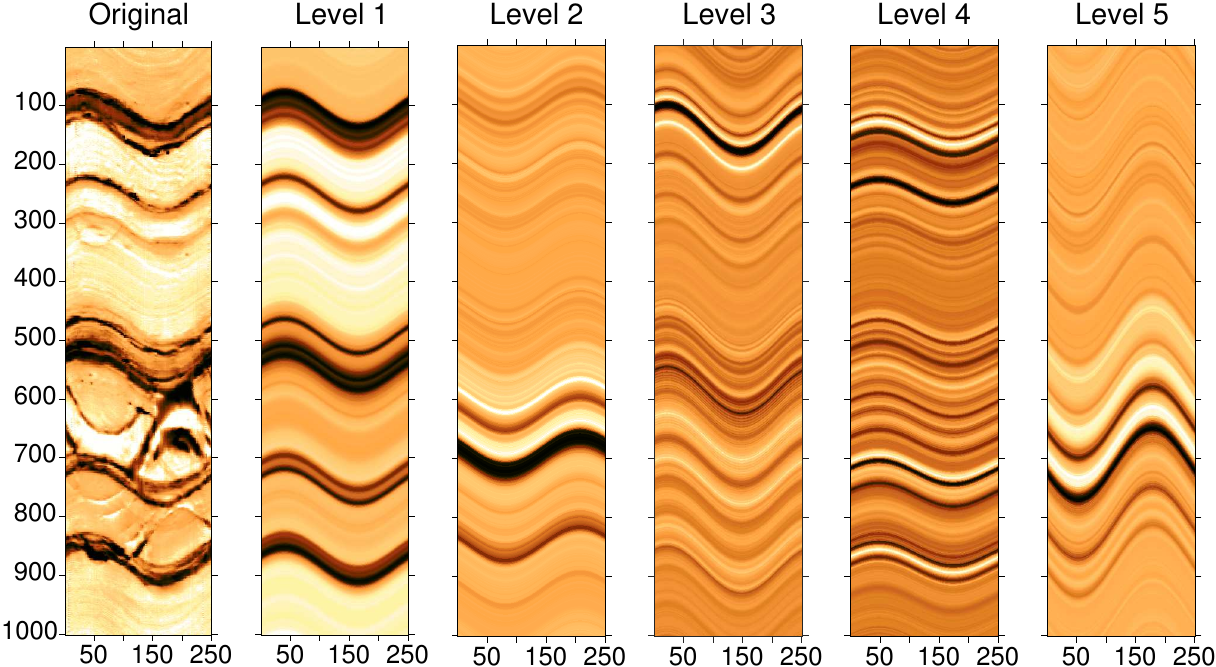}
\caption{Original image  and different levels $s^k$  ($k=1, \ldots, 5$) of the decomposition for case Fig.~\ref{fig:Cases}(b).  The values of $s^k$ are rescaled to the standard maximum and minimum}
\label{fig:level}       % Give a unique label
\end{figure}

\begin{figure}
 \centering
 \includegraphics[width=0.95\textwidth]{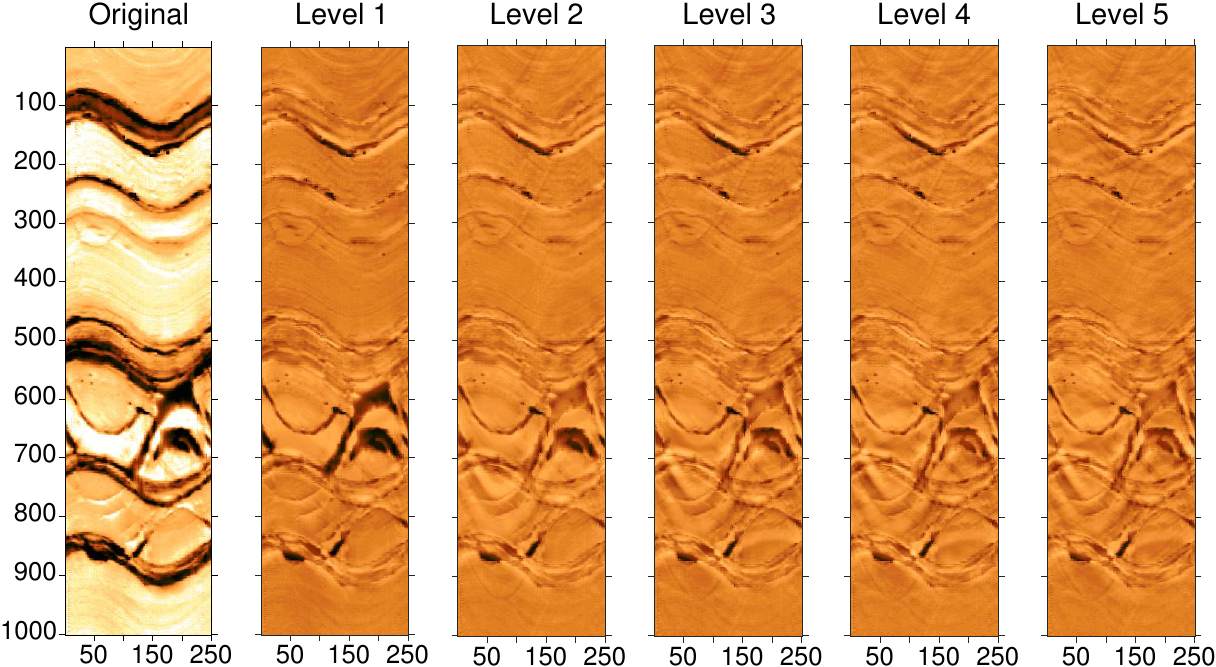}
\caption{Original image and residuals $res^k$ ($k=1, \ldots, 5$) after extraction of the approximations. The values of $res^k$ are rescaled to the standard maximum and minimum}
\label{fig:residuals}       % Give a unique label
\end{figure}

The first level $s^1$ in Fig.~\ref{fig:Cases} (the second column) dominates in the decomposition. Fig.~\ref{fig:level} concerns case Fig.~\ref{fig:Cases}(b). It represents the information brought from each level separately for five levels.   In this case, the variance of $s^k$ for $k>1$ is much less than the variance of $s^1$. The residuals after subtraction of the approximations of various levels for the same real-life case are presented in Fig.~\ref{fig:residuals}. The variance of the residual decreases at each step. The final residual is used for generation of the residual texture. It is clear from  Fig.~\ref{fig:residuals} that in this example the residual does not change significantly after the third step. 

To compute a value inside the volume, it is sufficient to identify the appropriate slices to which the point belongs at each level of this decomposition and then to compute the value of the residual texture inside the volume. For this purpose,  modelling of residual texture as a 3Dl moving average field \citep{francos1998,ojeda2010} is employed with calibrating the results to the observed statistics on the boundary of the borehole. Autocorrelations of this field are evaluated by its boundary values and are used for continuation of the field inside the volume. 

\begin{figure}
\centering
 \includegraphics[width=0.5\textwidth]{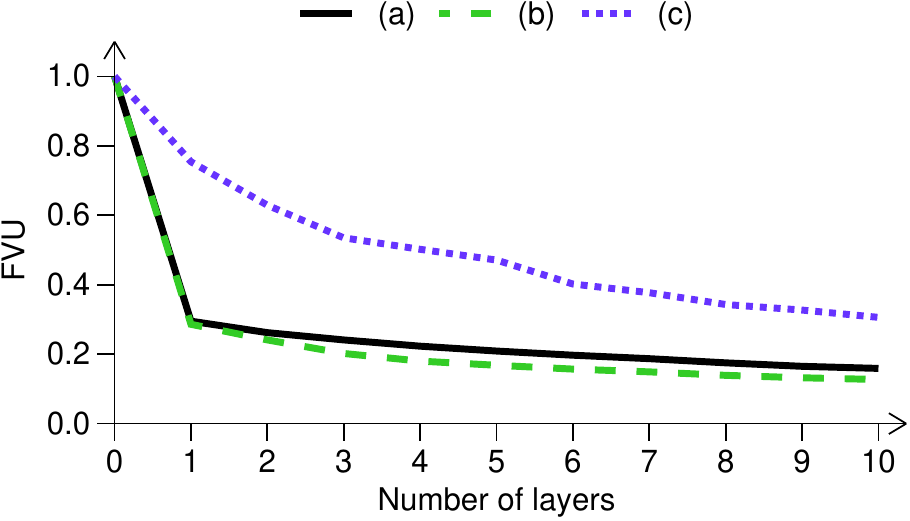}
\caption{FUV plots for the images presented in Fig.~\ref{fig:Cases} as functions of  number of slices used}
\label{fig:FUVplots}       % Give a unique label
\end{figure}

The problems of selection the best number of boundary elements is very similar to the problem of selection of the number of principal components in the signal approximation and there are several useful heuristics for its solution \cite{Cangelosi2007}. Usually, the first 3-5 boundary elements approximate the field quite well and the rest is left for the random field of texture. In Fig.~\ref{fig:FUVplots} the Fraction of Variance Unexplained (FUV) plots are presented as the functions of number of layers for the images used in the case study (Fig.~ \ref{fig:Cases}).  The variance of residual monotonically decreases but this decrease could be  slow  (Fig.~\ref{fig:residuals} where the difference between level 4 and level 5 residuals is not obvious).

This version of our algorithm seeks planar dominant features; but in principle, other geological features might be identified. Subtracting these from the original data leaves the subsidiary (i.e., less geologically significant) features which are isolated by an iteration of the method. Each set of subsidiary features is subtracted from the remaining data, and the process is repeated to isolate all of the significant subsidiary sets. A quality of fit test is performed on each slice family as it is identified. Examples of the sum of two-, and three- and four-slice  families are presented in the corresponding columns of Fig.~\ref{fig:Cases}.

To create the synthesized volume representing the rock removed during creation of the borehole, the main function, the $2,\ldots,L$ subsidiary functions, and the residual texture are summed in accordance with the expression
\begin{equation}\label{eq:1}
\hat{f}_{ij}  =\sum_{l=1}^L{s_{ij}^l +t_{ij}}.
\end{equation}
This equation is termed the reconstruction algorithm and is exact on the boundary. Moreover, for an arbitrary internal point of the cylinder $p=(x,y,z)$, the angle $\alpha$ is calculated as
\begin{equation}
\alpha=\arctan⁡(y/x)
\end{equation}
(that being an analogue in $\alpha_j$ in Fig.~\ref{fig:coordinate}(a)) and then the number of the slice in each slice family $k$ is calculated as
\begin{equation}
\nu^k=\left[\frac{R\sin\phi^k\cos(\psi^k-\alpha)+z\cos⁡\phi^k}{h}\right],
\end{equation}
where $\phi^k$ and $\psi^k$ are the angles that specify slice family at level $k$. Then  Eq.~(\ref{eq:1}) is used  to calculate the value of the field at this point.

The model matches the empirical boundary values exactly and preserves the statistical properties of the residual texture inside the volume. This internal residual texture is referred to as $t$. If necessary, this model can be improved by a multidimensional autoregressive moving average approach or other texture analysis algorithm.  There are many geological events which could be classified as texture. Most of them have specific geometries and may require specific tools for modelling. All these algorithms may be applied after extraction of the planar layer structure. 

The simple and universal moving average (MA) field is used for the modeling of residuals. Let the residual field   $res_{ij}= res_{ij}^k$  (\ref{res^k}) on the cylindric surface be given. A MA field $T(x,y,z)$ is constructed by averaging of 3D white noise in a sliding window. The sliding window is selected in the form of a rectangular parallelepiped in  such a way that the horizontal and vertical correlation radii of  $T(x,y,z)$ coincide with the correlation radii of residual $res_{ij}$ evaluated on the surface. After finding the dominant dip for the positive squared residual field $res_{ij}^2$, the field $res_{ij}$ is considered in the corresponding slices $\nu$.   The mean $\mu(\nu)$ and variance $\sigma^2(\nu)$ of $res_{ij}$ are evaluated in each slice  $\nu$. The values of $T(x,y,z)$ are shifted and scaled in each slice $\nu$  to have the same mean  $\mu(\nu)$ and variance $\sigma^2(\nu)$. The result is used for the inpainting of the texture.

This approach works satisfactorily if the residuals $res_{ij}^k$ are small enough with respect to the layer structure (Fig.~\ref{fig:residuals}). If it is necessary, the model could be refined by more advanced MA and autoregressive models \citep{francos1998, ojeda2010} or by other universal approaches, which include wavelet analysis and its various generalizations \citep{Portilla2000}, adapted kernel regression methods \citep{zhang2016} and methods of statistical physics \citep{wang2017} among others.   If the residual is not small then the (multi)layer structure $\sum_{l=1}^L s_{ij}^l$ is not dominant in Eq.~(\ref{eq:1}) and  special tools for the recognition and modelling of the geological structures may be needed.

\begin{figure}
\centering
  \includegraphics[width=0.5\textwidth]{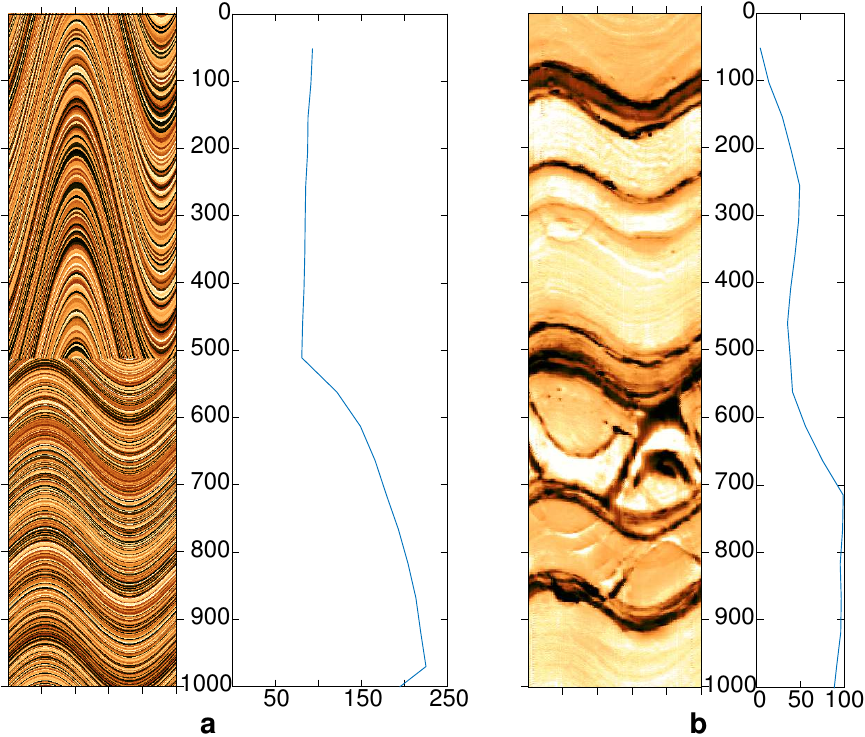} 
\caption{Dependences of the variance of residual (\ref{eq:fit}) (ordinate) on the depth  of a window  $z$ (abscissa) for the dominant dip: artificial combination ({\bf a}) and a real case Fig.~\ref{fig:Cases}(b) ({\bf b})}
\label{fig:S}       % Give a unique label
\end{figure}

The orientation of the structural or sedimentary structures may be not constant over the window; in addition the well path can also vary, thus one cannot expect that the sinusoids for a sufficiently long borehole are strictly parallel. If the change of direction in the sliding window is significant then a special {\it segmentation procedure} is needed. Fix a sliding window of the total depth $d$, with the internal coordinate $z\in [0,d]$. Find the optimal dominant dip in a shorter sliding window $[0,z]$ and calculate the variance of residual (\ref{eq:fit}) for this optimal approximation as a function of $z$: $v(z)$. If this function does not vary from a constant beyond a preselected {\it tolerance level}  $\varepsilon$ then    no segmentation is needed. If $v(z)$ deviates from the constant average value by more than $\varepsilon$ and a significant slope appears then segmentation is necessary and the  point of segmentation is the break point of the optimal piece-wise linear approximation of  $v(y)$. For the detailed description of segmentation algorithms  refer to the classical review by \citet{Keogh2004}. Examples of the function  $v(y)$ for the borehole images are presented in Fig.~\ref{fig:S} for two non-uniform images with segment infrastructure: artificial combination of two segments  (Fig.~\ref{fig:S}(a)) and a real example with variable structure (Fig.~\ref{fig:S}(b)) taken from Fig.~\ref{fig:Cases}(b) . The segmentation algorithm utilises  same calculation of residuals and search of optimal dips  as the basic approximation algorithm does.

Selection of parallel planar slices for the boundary elements formalizes the idea about planar geological structures. The choice of slices can be modified. For example, instead of parallel planes  the families of planes can be considered, which include a given straight line outside the cylinder. In this case, instead of the 2D optimization problem for function (\ref{eq:fit}) on a hemisphere of normal vectors  $\vec{n}$ one has to solve the four-dimensional optimization problem in the space of straight lines.  Algebraic surfaces provide many other possible choices of the families of slices but introduction of each sophisticated construction needs reasonable geological justification.

\section{Results}
\label{sec:2}

Figure~\ref{fig:Cases} shows three examples of surface reconstruction from initial images with different circumferential coverage states. The starting point for the upper set of images is image~\ref{fig:Cases}a with 65\% coverage typical of a well drilled with a 203-mm (8-inch) bit. Successive slice families are shown in images \ref{fig:Cases}b to \ref{fig:Cases}e, and image ~\ref{fig:Cases}f is the reconstruction formed from the sum of slice families plus residual texture. The reconstructed image has all of the main geological features of the original and has additionally filled the gaps present in the initial image. Filling the gaps in this way has  introduced some noise. The middle set of images is from another part of the same well, but this time the initial image \ref{fig:Cases}a has been inpainted with the morphological components analysis method of \citet{assous2014}, which reconstructs data missing in the gaps without introducing noticeable noise. The lower set of images is from a well drilled with a smaller bit and the circumferential coverage is 100\%. This example includes many fine fractures in addition to the crossing of planar bed boundaries.

\begin{figure}
\centering
  \includegraphics[width=0.45\textwidth]{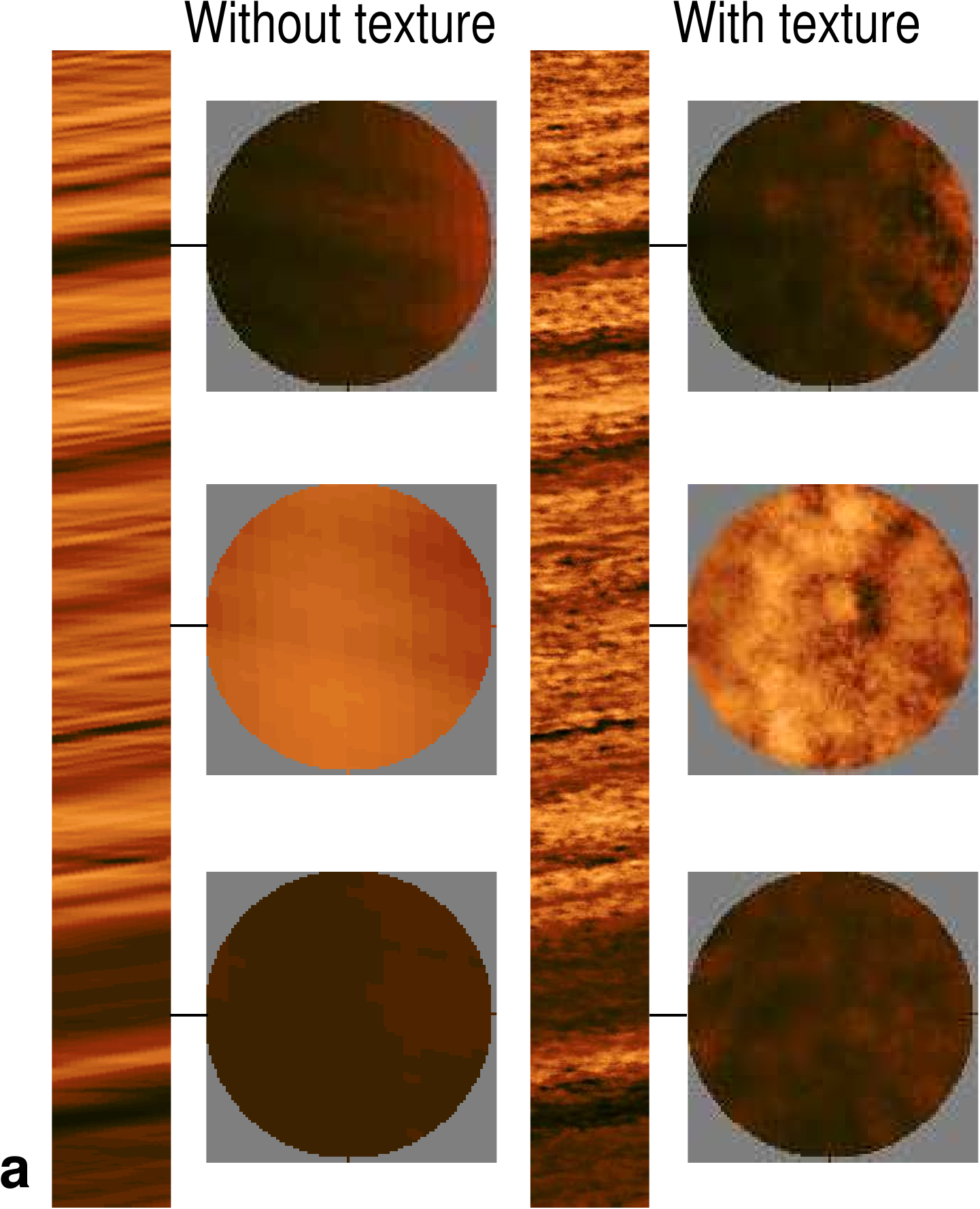} \hspace{3mm}
    \includegraphics[width=0.45\textwidth]{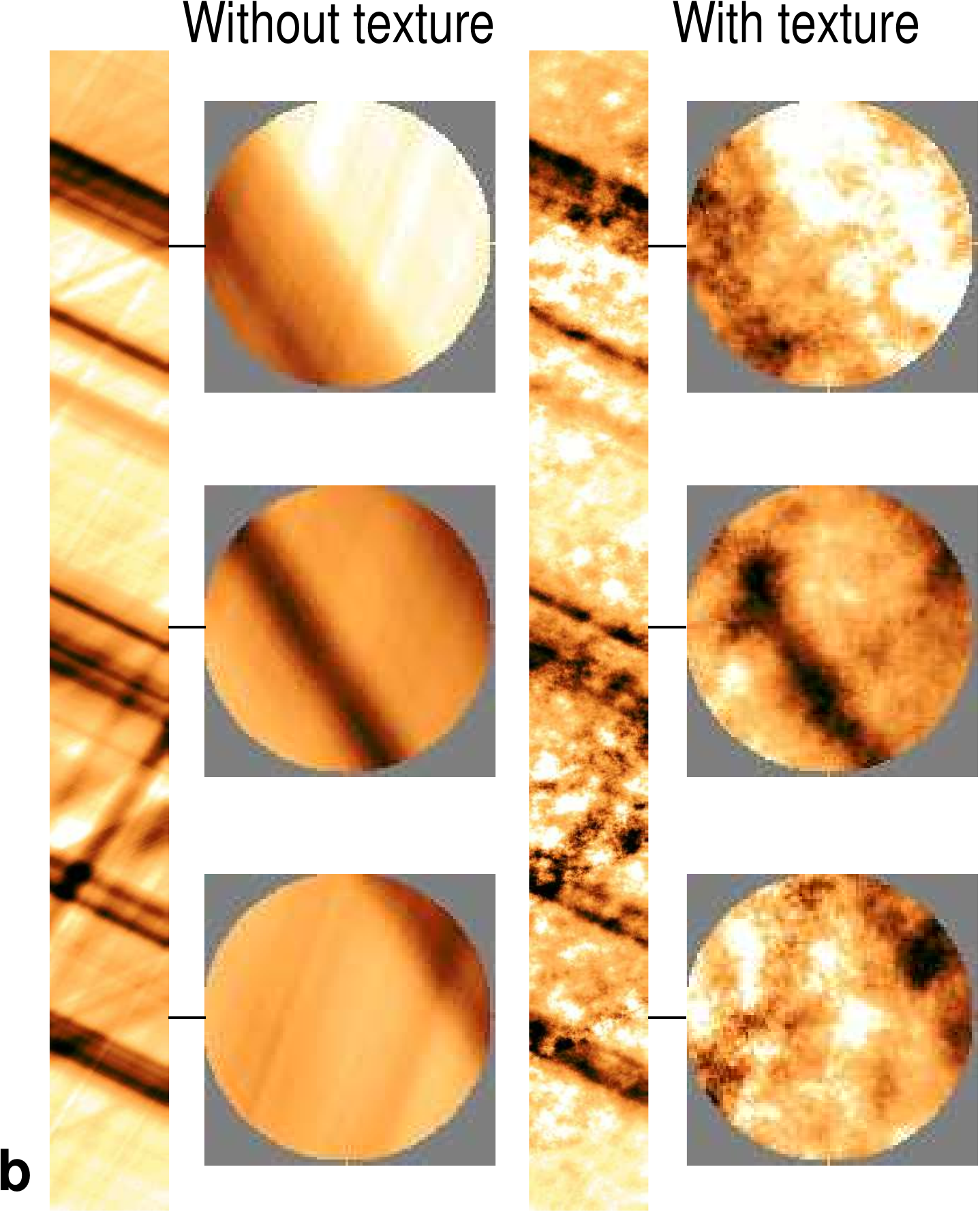}
      \includegraphics[width=0.45\textwidth]{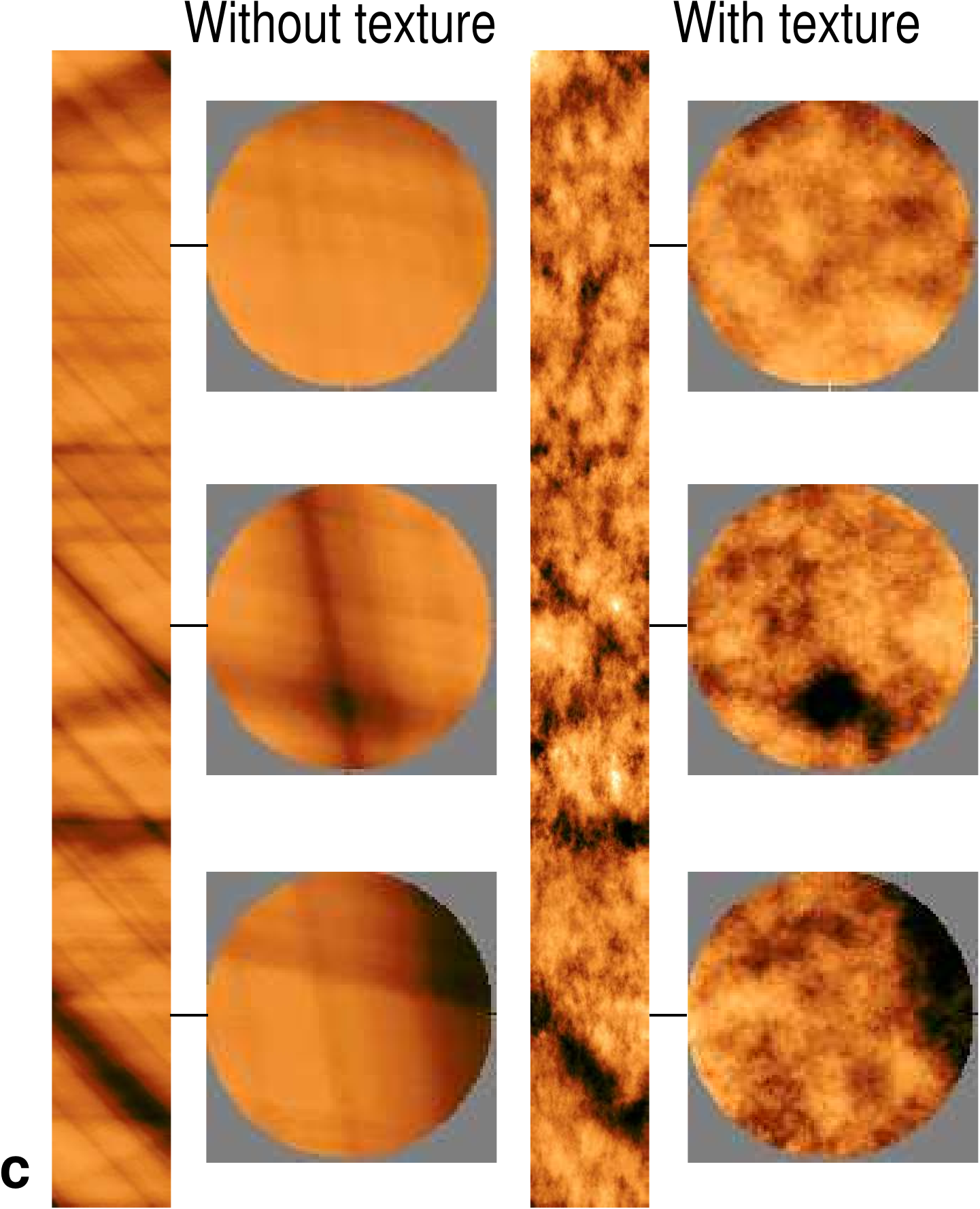}
\caption{Vertical cross sections and horizontal cross sections at three different depths reconstructed from the images in Fig.~\ref{fig:Cases}(a) -- \ref{fig:Cases}(c) in the same  order, with and without texture}
\label{fig:cross sections}       % Give a unique label
\end{figure}

Figure~\ref{fig:cross sections} shows horizontal and vertical slices through the solid images reconstructed from the images in Fig.~\ref{fig:Cases}. For each case, inpainted  images are without texture and images with texture are presented.

\begin{figure}
\centering
  \includegraphics[width=0.5\textwidth]{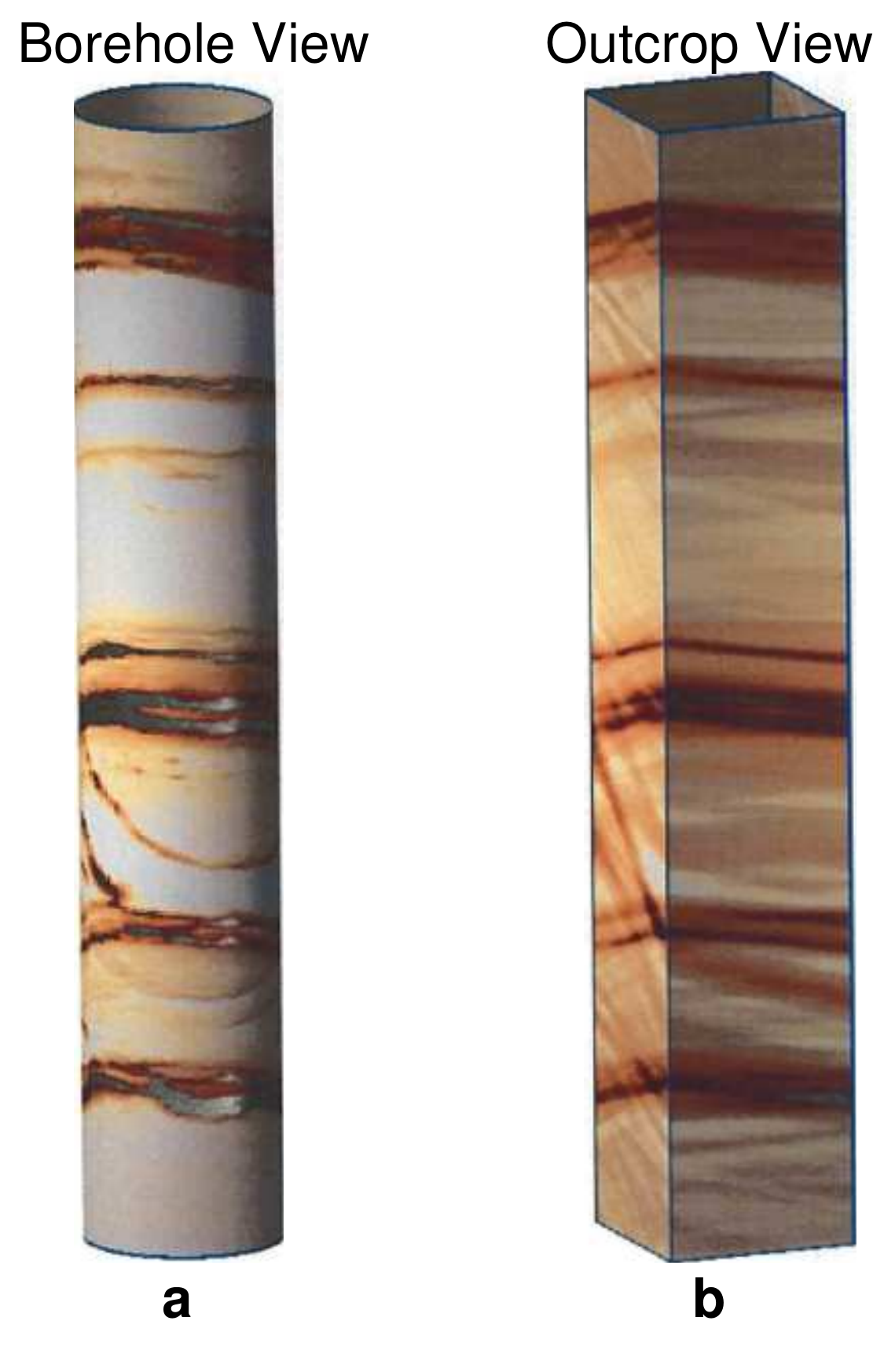}
\caption{Initial 2D inpainted circumferential image ({\bf a}) and 3D volume with two orthogonal slices ({\bf b})}
\label{fig:OrthoPainted}       % Give a unique label
\end{figure}

Figure~\ref{fig:OrthoPainted} shows an initial image with surface inpainting displayed in the conventional cylinder view and also a solid-volume view created by taking two orthogonal slices. Note that the planar geological features appear as curves on the conventional view, and as planes in the solid volume view. Relative to the conventional view, therefore, the angular relationships between geological features in the solid volume view are more straightforward for nonexpert interpreters to understand.

\begin{figure}
\centering
  \includegraphics[width=0.5\textwidth]{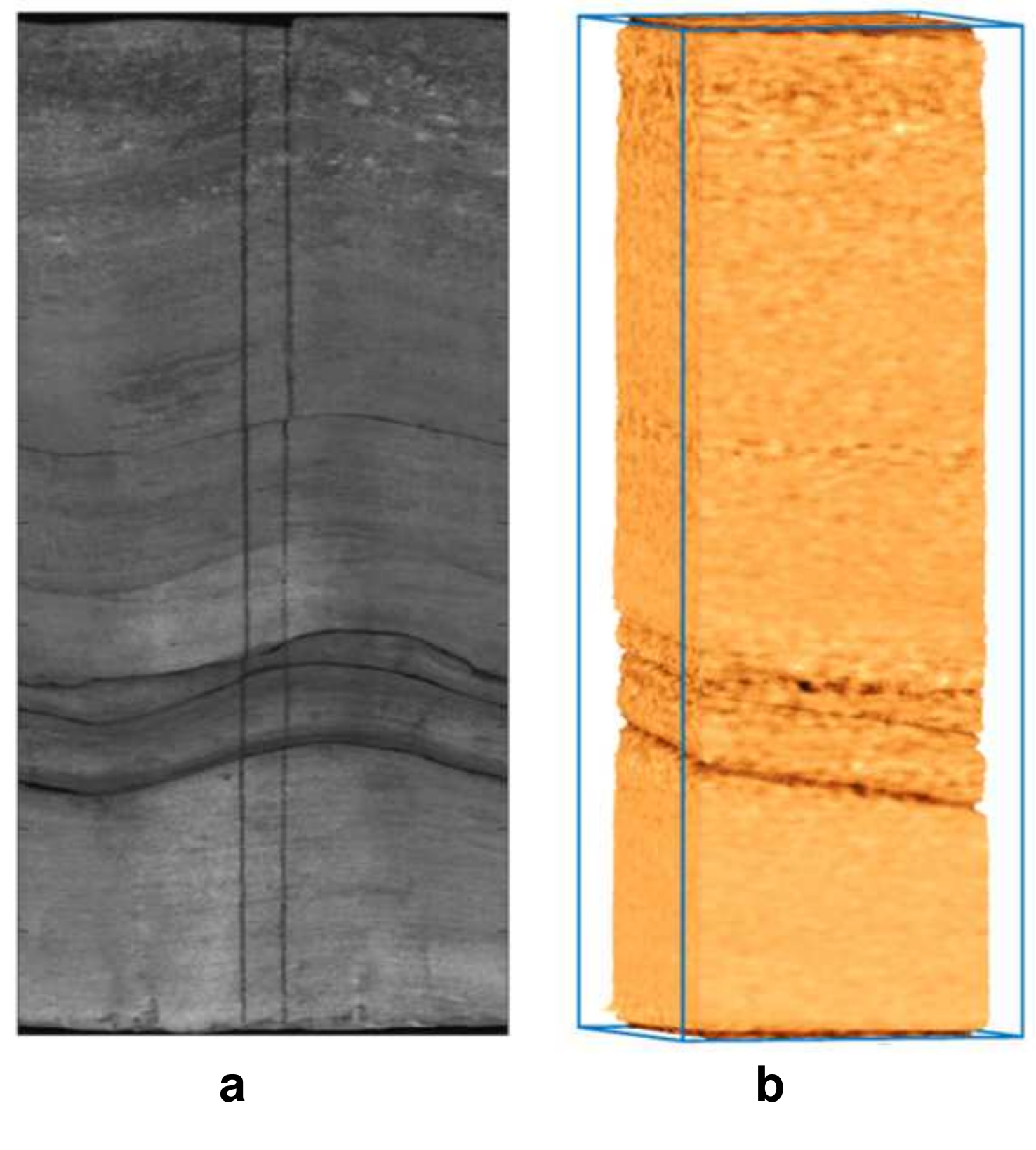}
\caption{Fullbore core surface scan ({\bf a}) and pseudo-outcrop visualization ({\bf b})}
\label{fig:Outcrop}       % Give a unique label
\end{figure}

Finally, the method has been applied to circumferential scans of fullbore core (Fig.~\ref{fig:Outcrop}). In this case, the external surface of solid volume has been modulated with a function of the grey level as a proxy for grain size (an indicator of depositional environment in sedimentary rocks). In the case of microresistivity images, a modulation based on resistivity value can be used.

\section{Conclusions}
\label{sec:3}

Image volumes created from circumferential microresistivity borehole images and optical core scans have been used to create pseudo-rock outcrop visualizations to help non-expert interpreters understand the earth-frame angular relationships between geological features. Modulating the external dimensions of the volume with a proxy for grain size adds information indicating depositional environment. The method focusses on planar geological features and textures, but it can be developed to include more complex structures. The method has been demonstrated on images with 65\% circumferential coverage typical of images acquired with pad-based sensors.

\begin{acknowledgement}
The authors acknowledge the help of John Winship, Technical Advisor at Weatherford Laboratories, UK, for providing core scans.
\end{acknowledgement}

\end{document}